%% file: ms.tex
\documentclass[apj]{emulateapj}
\usepackage{apjfonts}

\slugcomment{To appear in ApJ}
\received{May 24, 2006}
\revised{June 19, 2006}
\accepted{June 21, 2006}

\makeatletter
\def\ale{\mathrel{\mathpalette\gl@align<}}
\def\age{\mathrel{\mathpalette\gl@align>}}
\def\gl@align#1#2{\lower.6ex\vbox{\baselineskip\z@skip\lineskip\z@
\ialign{$\m@th#1\hfil##\hfil$\crcr#2\crcr\sim\crcr}}}
\makeatletter

\shorttitle{Spitzer/MIPS Imaging of NGC 650} 
\shortauthors{Ueta}

\begin{document}
 
\title{Spitzer/MIPS Imaging of NGC 650: \\
Probing the History of Mass Loss
on the Asymptotic Giant Branch} 

\author{%
Toshiya Ueta\altaffilmark{1}}

\affil{NASA Ames Research Center/USRA SOFIA Office,
Mail Stop 211-3,
Moffett Field, CA 94035, USA}
\email{tueta@sofia.usra.edu}

\altaffiltext{1}{NRC Research Associate/NASA Postdoctoral Research Fellow}

\begin{abstract}
We present the far-infrared (IR) maps of a bipolar planetary nebula
 (PN), NGC 650, at 24, 70, and $160\micron$ taken with the Multiband
 Imaging Photometer for Spitzer (MIPS) on-board the Spitzer Space
 Telescope.   
While the two-peak emission structure seen in all MIPS bands suggests
 the presence of a near edge-on dusty torus, the distinct emission
 structure between the $24\micron$ map and the 70/$160\micron$ maps
 indicates the presence of two distinct emission components in the
 central torus.  
Based on the spatial correlation of these two far-IR emission 
 components with respect to various optical line emission, we conclude
 that the $24\micron$ emission is largely due to the [\ion{O}{4}] line
 at $25.9\micron$ arising from highly ionized regions behind the
 ionization front, whereas the 70 and $160\micron$ emission is due to
 dust continuum arising from low-temperature dust in the remnant
 asymptotic giant branch (AGB) wind shell. 
The far-IR nebula structure also suggests that the enhancement of mass loss
 at the end of the AGB phase has occurred isotropically, but has ensued
 only in the equatorial directions while ceasing in the polar directions. 
The present data also show evidence for the prolate spheroidal
 distribution of matter in this bipolar PN.
The AGB mass loss history reconstructed in this PN is thus consistent
 with what has been previously proposed based on the past optical and
 mid-IR imaging surveys of the post-AGB shells.
\end{abstract}


\section{Introduction}

NGC 650 (PK 130-10\fdg1, M 76, Little Dumbbell Nebula) is a large
($\sim300\arcsec$; \citealt{balick92}) bipolar planetary nebula 
(PN) of the ``late butterfly'' type \citep{balick87}.
The nebula structure in the optical consists of the bright rectangular
core of $95\arcsec \times 40\arcsec$ (the long side perpendicular to the
bipolar axis) and a pair of fainter lobes extending $\sim90\arcsec$ and
$\sim150\arcsec$ from the central star that is attached to the
long side of the rectangular core.
The central core was long suspected to be a nearly edge-on torus,
and the most recent kinematical study has eloquently demonstrated that
the core is an inclined torus and the lobes are blown-bubbles expanding
into the polar directions \citep{bryce96}.  

Under the framework of the widely-accepted general interacting stellar
wind (GISW) model (e.g., \citealt{balick87}), the formation of bipolar
PNs is understood as a two-step process.
First, the progenitor star has to lose its envelope material via mass
loss (${\dot M} \sim 10^{-6}$ M$_{\odot}$, $v \sim 10$ km s$^{-1}$)
during the asymptotic giant branch (AGB) phase of the evolution.
This AGB wind leads up to the so-called superwind
(${\dot M} \sim 10^{-4}$ M$_{\odot}$, $v \sim 20$ km s$^{-1}$) at the
end of the AGB phase prior to the exhaustion of the envelope
material. 
During the superwind phase, mass loss is expected to cause concentration 
of the ejected matter into the equatorial plane. 
The AGB mass loss therefore results in the equatorially-enhanced
circumstellar shell.

Then, a hot and tenuous fast wind (${\dot M} \sim 10^{-9}$ M$_{\odot}$,
$v \sim 10^{3}$ km s$^{-1}$) begins to emanate from the progenitor just 
prior to the beginning of the PN phase when the circumstellar material
starts to ionize.
This fast wind pushes into the surrounding equatorially-concentrated 
envelope, and the flow of the wind is thus channeled into the polar
directions.
The preferential wind flow towards the polar directions creates the
typical bipolar lobes as wind-blown bubbles.
The GISW scheme is thus capable of producing various PN shapes
(both bipolar and spheroidal shells) depending primarily on the degree
of the equatorial enhancement in the surrounding AGB shell. 

Hence, in order for the GISW scheme to work, the equatorially-enhanced
AGB shell must be present when a fast wind is initiated.
The GISW model itself, however, does not address how to generate the
equatorial enhancement in the AGB wind shell. 
The presence of such circumstellar shells with a built-in equatorial
density enhancement has been confirmed by imaging surveys of proto-PNs
in thermal IR dust emission from the innermost torus 
(e.g., \citealt{skinner94,meixner97,dayal98,meixner99,ueta01}) and
dust-scattered star light (e.g., \citealt{ueta00}). 
Based on the results from the largest mid-IR and optical imaging surveys 
to date \citep{meixner99,ueta00}, a so-called ``layered shell model'' has
been proposed to explain {\sl both} the bipolar and elliptical
proto-PN morphologies \citep{ueta02,ueta03}.
In this model, the proto-PN shells are thought to consist of three
generic layers, each of which possesses a specific structure reflecting
the geometry of mass loss at the time.
The outer spherically symmetric layer represents isotropic mass loss 
during the early AGB phase, while the inner toroidal layer embodies 
equatorially-enhanced mass loss during the superwind epoch at the end of 
the AGB phase. 
Between these layers there is an intermediate spheroidal layer, which  
results from a gradual transformation of the mass loss geometry from
isotropic to equatorially-enhanced over the course of the AGB phase.

The unique strength of this layered shell model is its versatility.
Radiative transfer calculations using this model have successfully
reproduced both bipolar and elliptical proto-PN morphologies by simply
varying the degree of the equatorial density enhancement (i.e., optical
depth) of the shell \citep{meixner02,ueta03}.  
The implication of the layered shell model is that there is no
intrinsic distinction between the elliptical and bipolar proto-PN shells  
other than the degree of the equatorial density enhancement.
Bipolar PNs would emerge from equatorially-enhanced proto-PNs and 
elliptical PNs would descend from proto-PNs of rather isotropic density
distribution.  
The presence of the spheroidal intermediate layer is the key to account
for the elongation along the polar axis in proto-PN shells (e.g.,
\citealt{ueta01,meixner02,ueta03,meixner04}). 

It is clear that the equatorially-enhanced superwind {\sl initiates} the
subsequent aspherical structure development that eventually results in
the observed complex PN structures.  
However, it is still not yet evident how the shell structure development
ensues in the AGB and post-AGB winds.
How does the spheroidal density distribution arise prior to the
equatorial enhancement at the end of the AGB phase?
Is the post-AGB mass loss significant in shell shaping?
Does a fast wind simply ``snow-plow'' the surrounding material, tracing
the pre-existing toroidal density distribution, or contribute to an
additional equatorial enhancement? 
These are only a few questions concerning the shell structure
development in the circumstellar shells of evolved stars.

Further to understand the role of mass loss in the shell structure
development during the AGB phase and beyond, the remnant AGB shells need
to be investigated in PNs via sensitive far-IR observations of thermal
dust emission, since such remnant AGB shells are very much dispersed and
cold. 
In the following, we present the results of Spitzer Space Telescope
far-IR mapping observations of a bipolar PN, NGC 650, to probe the
history of AGB and post-AGB mass loss in this object.
Below, we describe the data set and reduction procedure (\S 2),  discuss
the results (\S 3), and summarize conclusions (\S 4).  

\section{Observations and Data Reduction}

NGC 650 was observed with the Multiband Imaging Photometer for Spitzer 
(MIPS; \citealt{mips}) on-board the Spitzer Space Telescope \citep{sst}
on 2004 September 18 as part of GTO first year observations (Program ID
77; AOR KEY 9548032).
The MIPS observations were performed in the scan-map mode using the
medium scan rate. 
The scan was done along four scan legs of $0\fdg5$ long each with the
cross-scan stepping of $148\arcsec$.
The whole scan sequence was repeated twice.
The entire three-band dataset was retrieved from the Spitzer Science
Archive using the Spitzer-Pride software tools (Spot/Leopard).

Data reduction was done mostly with the 2005 September 30 version of the
Mosaicker software provided by the Spitzer Science Center
(SSC)\footnote{\url{http://ssc.spitzer.caltech.edu/postbcd/}} and the
Ge Reprocessing Tools (GeRT; Ver.\ 041506 of S14
processing)\footnote{\url{http://ssc.spitzer.caltech.edu/mips/gert/}} .
The $24\micron$ map was constructed using the Mosaicker from the Basic
Calibrated Data (BCD) products that were produced from the raw data via
the SSC pipeline.
Upon making a mosaic map, anomalous pixels were removed by using the
dual outlier detection algorithm in the Mosaicker. 

Since the post-BCD mosaicked maps at 70 and $160\micron$ generated by
the SSC pipeline showed obvious artifacts, we generated these maps
directly from the raw data using the GeRT and the Mosaicker.
First, we corrected bad stimulator response calibration affected by the
bright objects (i.e., the target), in order to optimally remove the
time-dependent responsivity variation of the detector.
This is done by ignoring the stimulator flash frames on and near the
target. 
Second, we removed the ``streakings'' due to residual slow response
variations of the detector by applying a high-pass time filter.
Since we are interested in the extended far-IR structure of the object,
we did not perform column filtering for the $70\micron$ band data.
Instead, we used a relatively small median-count value of six when
applying a high-pass time filter.
Upon applying the time filter, we masked out the target to avoid
extended faint emission from being mistakenly filtered out.
These GeRT processes resulted in cleaner BCDs.
Then, we constructed mosaic maps from the GeRT-processed BCDs using 
the Mosaicker. 

The resulting maps are in $1\farcs28$ pix$^{-1}$, $4\farcs92$
pix$^{-1}$, and $7\farcs99$ pix$^{-1}$  for the 24, 70, and
$160\micron$ bands, respectively.
After making artifact-removed maps for all three bands, we subtracted
sky emission, which was estimated at off-source positions that are 
free from background sources. 
The sky emission was determined to be 20, 9, and 27 MJy sr$^{-1}$
respectively in the 24, 70, and $160\micron$ bands.
These values are consistent with the estimated values of the date of
observations obtained using the Spot software. 
In order to facilitate direct comparison of the nebula structure between
two different MIPS bands, the 24 and $70\micron$ images were convolved
with the beam of the longer wavelength bands and their pixel scales were
made to match with that of the longer wavelength band.
Flux calibration of the maps were done using the conversion factors
provided by the SSC.

\section{Results and Discussion}

\subsection{NGC 650 in the Far-IR}

\begin{figure*}
 \begin{center}
 \includegraphics[width=6in]{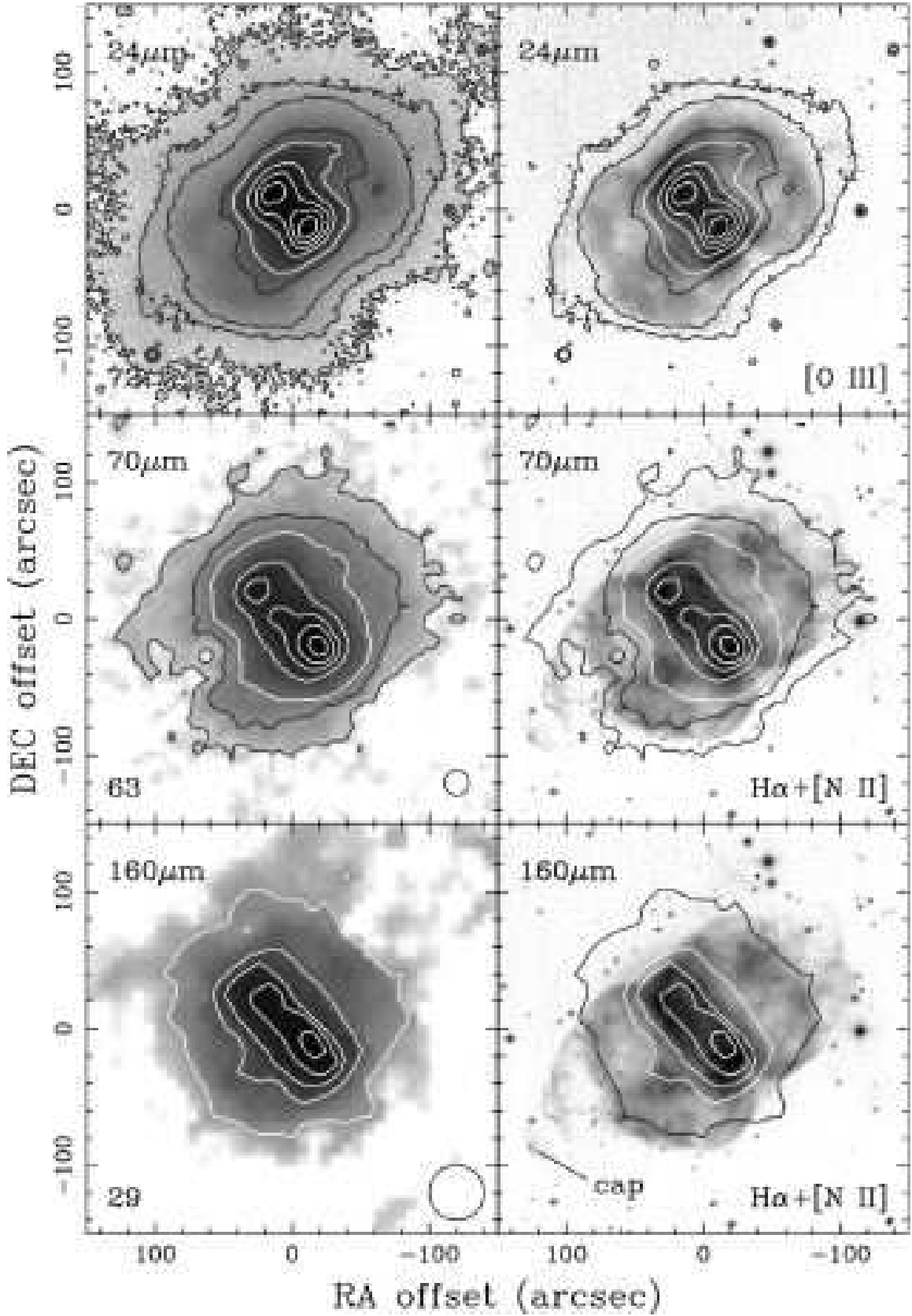}
 \end{center}
 \caption{\label{2x2map}%
 Grayscale MIPS mosaicked maps of NGC 650 at $24\micron$ (top left),
 $70\micron$ (middle left), and $160\micron$ (bottom left) overlaid with
 contours representing 90 to 10\% of the peak (with the 20\% interval;
 in white) and  5, 1, 0.5, and 0.15\% of the peak (in black).  
 Also presented are grayscale IAC optical maps in the [\ion{O}{3}] 5007 \AA
 band (top right) and in the H$\alpha$+[\ion{N}{2}] 6600 \AA band
 (middle and bottom right), overlaid with the same far-IR contours as
 left panels (except for the lowest in the $24\micron$ band for clarity).
 The images are in nominal orientation (N is up, E to the left) and
 centered at the position of the central star (RA [2000] =  
 1$^{\rm h}$42$^{\rm m}$19\fs656, Dec [2000] =
 51\arcdeg34\arcmin32\farcs79; \citealt{k98}).  
 The angular offsets in arcsec are shown by the tickmarks.
 A circle at the lower right corner in the left panels represents the
 beam size (FWHM) of each band ($6\arcsec$, $18\arcsec$, and $40\arcsec$
 for the 24, 70, and $160\micron$ bands, respectively). 
 The number shown at the lower left corner in the left panels is the
 peak intensity in MJy sr$^{-1}$.
 The lowest contour is the 3 $\sigma$ level (0.1, 0.9, 3 MJy sr$^{-1}$
 respectively for the 24, 70, and $160\micron$ bands).} 
\end{figure*}

Figure \ref{2x2map} presents the MIPS mosaicked maps of NGC 650 at
$24\micron$ (top left), $70\micron$ (middle left), and $160\micron$ 
(bottom left) in grayscale and contours. 
Also displayed for comparison are optical narrow-band images in the
[\ion{O}{3}] band at 5007 \AA (top right) and H$\alpha$+[\ion{N}{2}]
band at 6600 \AA (middle and bottom right) in grayscale, obtained from
the Instituto de Astrof\'{\i}sica de Canarias (IAC) Morphological
Catalog of Northern Galactic Planetary Nebulae
Archive.\footnote{\url{http://www.iac.es/nebu/nebula.html}.}
These optical images are overlaid with the far-IR contours.
The figure caption gives the complete detail of the images.

In all bands, the nebula core is resolved into two emission peaks that
delineate the bipolar waist of the optically bright core (i.e., the
short sides of the core).   
The southwestern peak is brighter than the northeastern peak in all
bands.
The intensity of the brighter (southwestern) peak is 72, 63, and 29 MJy
sr$^{-1}$ while that of the dimmer (northeastern) peak is 61, 35, and 23
MJy sr $^{-1}$, both respectively in the 24, 70, and $160\micron$ bands.
Previous kinematic studies of the nebula in the optical emission lines  
have shown that the bright core is an expanding torus (dubbed a ``napkin
ring'') inclined at about $75^{\circ}$ with respect to the line of sight
(e.g., \citealt{bryce96}).  
A similar two-peak emission structure at mid-IR has been commonly
observed in optically thin, equatorially-enhanced proto-PN dust shells
that are oriented near edge-on (e.g., 
\citealt{meixner93,skinner94,ueta01,gledhill03}).  
The resolved far-IR structure in this PN is consistent with the
mid-IR structure of progenitor objects, suggesting that the
innermost torus of warm dust in a proto-PN shell would keep its toroidal
shape into the PN phase while the dust temperature decreases upon expansion.
The two-peaked far-IR structure thus represents the limb-brightened
edges of a near edge-on optically thin dusty torus, at which the column
density of the far-IR emitting matter is the largest.

Far-IR emission in the nebula is more extended than the optically bright 
core region.
Emission is detected at least 10\% of the peak intensity level in the
regions interior to the optical inner lobes ($\ale90\arcsec$ from the
center) in all bands. 
Fainter emission appears to be more extended along the bipolar axis
beyond the optical inner lobes at shorter wavelengths.
In the $24\micron$ band, emission at $\sim0.5\%$ of the peak level 
($\sim 10\sigma$ detection) is observed in regions corresponding to the
outer lobes.  
There is a detached ``cap'' detected in the optical at the tip of the 
southeast lobe (marked in the bottom right panel of Fig.\ \ref{2x2map}).
\citet{hora04} have reported detection of emission from this cap in the
Spitzer/Infrared Array Camera (IRAC) bands (and even from the opposing
one in the northwest lobe). 
Although the $24\micron$ map shows very low-level ($\sim$ three $\sigma$)
emission in the regions corresponding to the southeast cap, it is
uncertain whether we have positive detection of the cap or not.
There appears to be an isolated patch of emission near the northwest
corner of the panel at ($-140\arcsec$, $120\arcsec$).
This is, however, not the opposing cap: it is simply a background source.

\begin{figure}
\begin{center}
 \plotone{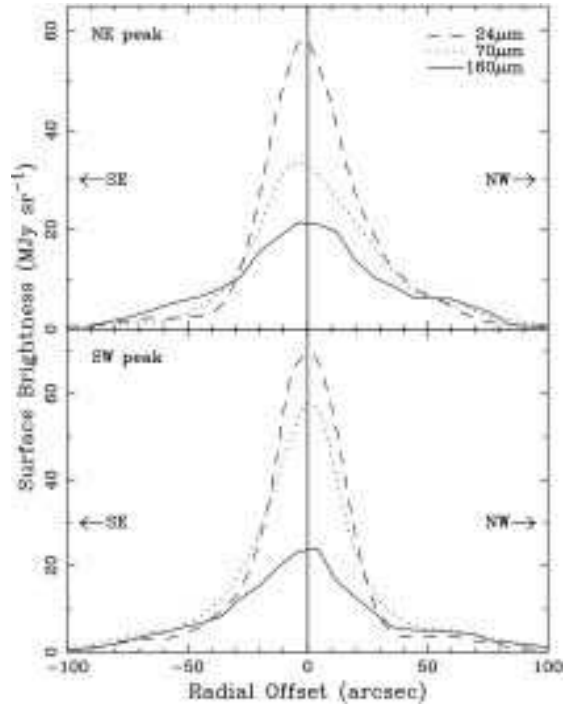} 
\end{center}
\caption{\label{peakcuts}%
 Surface brightness profiles of the emission peaks in the
 direction of the polar axis. The top (bottom) panel is for the
 northeast (southwest) peak. The cuts are made along lines of
 $25\arcsec$ width parallel to the bipolar axis (of $\mbox{PA} =
 -45^{\circ}$).  The 24, 70, and $160\micron$ band data are shown
 respectively by dotted, dashed, and solid lines. Positive offsets
 corresponds to the northwest direction as indicated in the panels.
 While the peak location is shifted towards
 southeast (northwest) in the northeast (southwest) peak, the peak
 itself is more elongated towards the opposite, northwest (southeast)
 direction.}  
\end{figure}

Emission of intermediate strengths ($\sim 1$ to $20\%$ of the peak) shows
an interesting morphology of the nebula.
In the optical image, there is a region of enhanced brightness in both
of the inner lobes.
The bright region in the optical northwest lobe seems to be connected to
the northeast end of the waist, while the bright region in the southeast
lobe appears to be attached to the southwest end of the waist.
On one hand, a similar trend is seen in the far-IR maps.
The northeast (southwest) peak is more elongated towards the northwest
(southeast) lobe, corresponding to the optically bright region.
On the other hand, the actual location of the northeast (southwest) peak
appears to be displaced towards the southeast (northwest) direction,
which is opposite to the direction of the peak elongation mentioned
above. 
Figure \ref{peakcuts} is presented to show this morphological trend by
surface brightness profiles.

While the overall boxy shape of the optically bright core is oriented
perpendicular to the bipolar axis, the emission peaks do not seem to be
aligned with the equatorial plane.
The elongation of the peaks is best seen in the $24\micron$ map,
while the peak displacement is well displayed by the $70\micron$ map. 
This morphological trend is consistent with that has been seen in the
optical (\citealt{balick87,bryce96}), H$_{2}$ \citep{kastner96}, and
IRAC \citep{hora04} bands, and gives the nebula point-symmetric appearance.   

Aperture photometry above the three $\sigma$ level (the lowest contour in
Figure \ref{2x2map}) yields flux measurements of $4.51\pm0.04$,
$6.04\pm0.30$, and $4.83\pm0.95$ Jy in the 24, 70, and $160\micron$
band, respectively.  
These values are listed in Table \ref{flux} together with the past
measurements taken by Infrared Astronomical Satellite (IRAS).
The measured flux at the $24\micron$ MIPS band is about $50\%$ higher
than that at the $25\micron$ IRAS band.
The color correction on the $25\micron$ IRAS flux does not seem to
account for the entire $50\%$ discrepancy.
Since the entire $24\micron$ MIPS band is overlapped with the
$25\micron$ IRAS band, the discrepancy cannot be attributed to the
presence of strong IR fine-structure lines in the $24\micron$ MIPS
band that are not covered by the $25\micron$ IRAS band.
Sensitivity of the $24\micron$ MIPS band is superior to that of the
$25\micron$ IRAS band, and the $24\micron$ MIPS map shows a very faint 
extended nebulosity. 
Thus, the discrepancy could be attributed to the IRAS measurements
missing the faint emission.
Aperture photometry in the $24\micron$ MIPS band using the estimated
measurement threshold at the $25\micron$ IRAS band yields 3.67 Jy.
Considering also the beam dilution effect expected in the IRAS data, we
conclude that the $50\%$ flux discrepancy is due to faint emission in
the extended nebulosity that was not previously detected by IRAS.

\input{tab1.tex}

\subsection{Possible Two-Component Structure in the Central Torus}

\begin{figure}
\begin{center}
 \plotone{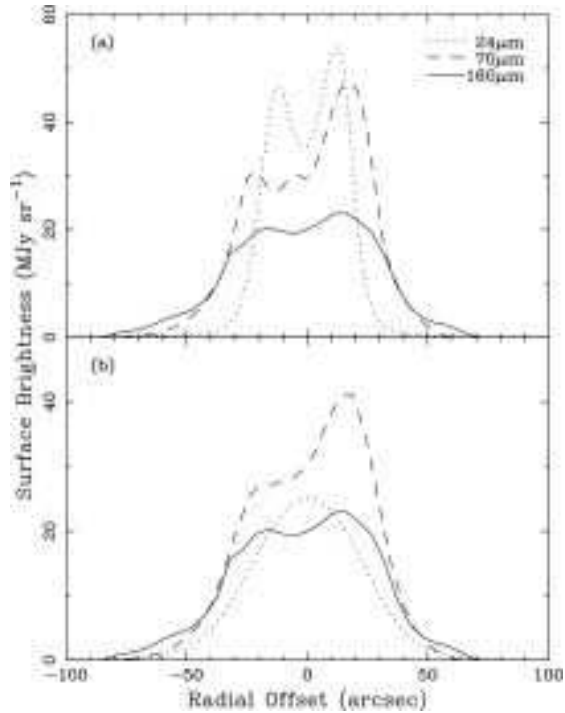} 
\end{center}
\figcaption{\label{eqcuts}%
 Surface brightness profiles of the emission peaks along the equatorial
 plane: (a) profiles at the resolution of each band (FWHM of $6\arcsec$, 
 $18\arcsec$, and $40\arcsec$ for the 24, 70, and $160\micron$ bands,
 respectively) and (b) (convolved) profiles at the resolution of the
 $160\micron$ band. The cuts are made along lines of $40\arcsec$ width
 perpendicular to the bipolar axis (of $\mbox{PA} = 45^{\circ}$).
 Positive offsets corresponds to the southwest side of the nebula. 
 While the 70 and $160\micron$ peaks arise from the same locations, the
 $24\micron$ peaks arise from spatially distinct locations closer to the
 center.} 
\end{figure}

In this section we give a more detailed analysis of the two-peak
structure of the central torus.
The measured peak separation is $36\farcs7$, $58\farcs5$, and
$42\farcs3$ respectively in the 24, 70, and $160\micron$ bands. 
Figure \ref{eqcuts}a displays surface brightness profiles along the 
equatorial plane taken over the whole width ($40\arcsec$) of the
rectangular optical core.

If the equatorial density profile follows a power law with the maximum
density at the inner radius (i.e., $\rho = \rho_{\rm in} r^{-\alpha}$,
where $\rho_{\rm in}$ is the density at the inner radius, $r_{\rm in}$),
the largest column density of the far-IR emitting material along the
line of sight occurs near the inner radius (at locations $r_{\rm
in}$ away from the center along the equatorial plane).  
This means that the far-IR peaks occur near the inner radius when
the dusty circumstellar shell is optically thin to far-IR radiation, as
has been shown in dust radiative transfer models for proto-PN dust
shells (e.g., \citealt{ueta01,meixner02,ueta03,meixner04}). 

When the spatial resolution of the imaging bands is taken into account,
the peaks are expected to occur progressively closer in the bands at
longer wavelengths until the peaks become unresolved. 
This is why the $70\micron$ peak separation is larger than the
$160\micron$ peak separation.
In fact, when the $70\micron$ image is convolved to have the spatial
resolution of the $160\micron$ band, the 70 and $160\micron$ peaks occur
at identical locations as demonstrated in Figure \ref{eqcuts}b.
Given the spatial coincidence of the emission peaks in the 70 and
$160\micron$ bands, we interpret that these peaks represent the
limb-brightened edges of a dusty torus.
In this picture, the peak separation at the 70 and $160\micron$ bands
corresponds to the inner diameter of the dusty torus at a relatively low
temperature. 

On the contrary, the $24\micron$ peak separation does not follow this
trend.  
If the peaks in the $24\micron$ band represented the Wien head of the
spectral energy distribution (SED) of the dusty torus seen in the 70 and
$160\micron$ bands, the peaks should have been well resolved with the
peak separation slightly larger than that in the $70\micron$ band.
However, this is not the case.
Similarly, the 70 and $160\micron$ peaks cannot be the Rayleigh-Jeans
tail of the SED of a dusty torus whose presence can be inferred from the 
$24\micron$ map.
Since the beam size of the $160\micron$ band is $40\arcsec$, the
$24\micron$ peaks ($36\farcs7$ separation) cannot be resolved.
The $24\micron$ band profile convolved with the $40\arcsec$ beam
indeed shows only one peak as demonstrated in Figure \ref{eqcuts}b.
Hence, while the $24\micron$ two-peak structure also suggests the
presence of an edge-on torus, this torus must possess distinct
physical properties with respect to the torus seen in the 70 and
$160\micron$ bands.  

Imaging in various optical emission lines have suggested that ionization
fronts are found only in very selected regions in the core and that the
lobe edges are not ionization fronts \citep{balick87,balick92}. 
In particular, \citet{balick87} presented an [\ion{N}{2}] image showing
two emission peaks about $70\arcsec$ apart in the core and a \ion{He}{2}
image displaying two peaks about $50\arcsec$ apart.
The [\ion{N}{2}] and \ion{He}{2} peaks respectively represent ``low''
and ``high'' ionization regions. 
The [\ion{N}{2}] peak separation is about $45\arcsec$ if convolved with
the $40\arcsec$ beam, and thus, these peaks spatially corresponds to the
possible low-temperature dusty torus revealed by the $70$ and
$160\micron$ maps.   
Similarly, the \ion{He}{2} peak separation is about $40\arcsec$ if
convolved with the $6\arcsec$ beam of the MIPS $24\micron$ band, and
thus, the \ion{He}{2} peaks spatially corresponds to the other torus
revealed by the $24\micron$ maps.  
This is also demonstrated by the IAC optical images in
Figure \ref{2x2map} (and Figure \ref{2x1map} for a close-up of the
equatorial region).  
The ``high'' ionization region represented by [\ion{O}{3}] emission
corresponds to the $24\micron$ peaks, while the ``low'' ionization
region delineated by H$\alpha$ and [\ion{N}{2}] emission is coincident
with the 70 and $160\micron$ emission distribution.

In fact, \cite{lp87} have shown that the mid- and far-IR structures of
the Helix nebula (NGC 7293) are markedly distinct in the IRAS data, in
which emission in the mid-IR ($\le 25\micron$) is concentrated in the
central region.
These authors have attributed the $25\micron$ band emission mostly to
the [\ion{O}{4}] line at $25.9\micron$. 
The following investigations have shown that the [\ion{O}{4}]
distribution is also co-spatial with the \ion{He}{2} distribution in the
nebula \citep{odell98,speck02}.
Moreover, H$_{2}$ emission, which suggests the presence of a high
density molecular reservoir, arises from the limb of the bipolar waist,
which is spatially coincident with the 70 and $160\micron$ peaks
\citep{kastner96,hora04}. 

Therefore, we conclude that the far-IR toroidal structure in
NGC 650 consists of two distinct components.
The first component is represented by the $24\micron$ peaks and is
a high-temperature torus encompassing the regions of high ionization.
Given the spatial correlation between the $24\micron$ emission region
and the emission regions of high ionization species and similarities
between NGC 650 and the Helix , we also suggest that the detected
$24\micron$ emission is largely due to the [\ion{O}{4}] line at $25.9\micron$. 
The second component is represented by the 70 and $160\micron$ peaks and 
is a low-temperature dusty torus that is exterior to the
high-temperature torus and is not yet very much ionized by the radiation
from the central star. 

\subsection{The Remnant AGB Wind Shell}

In this section, we continue our discussion on the low-temperature outer
torus revealed in the 70 and $160\micron$ bands.
As reviewed in \S 1, the circumstellar dust shell of post-AGB stars
develops its toroidal form in the innermost regions during the superwind
phase (e.g., \citealt{meixner99,ueta00}).
Such a torus would then keep expanding while decreasing its temperature
during the subsequent evolutionary phases into the PN phase.  
Hence, we postulate that the dust torus seen in the 70 and $160\micron$
maps is the remnant AGB wind shell that has been expanding since it was
generated by the superwind at the end of the AGB phase.  

Given that this outer dust torus is the remnant AGB wind shell, its
surface brightness profiles would yield 
valuable information pertaining to the AGB mass loss history.  
If the farthest extent of the highly ionized region has reached only up
to where the $24\micron$ peaks are, the surface brightness distribution
beyond the $24\micron$ peaks reflects the mostly pristine AGB mass loss
history that is still relatively undisturbed by the passage of the fast
wind and ionization front. 

\begin{figure}
\begin{center}
 \plotone{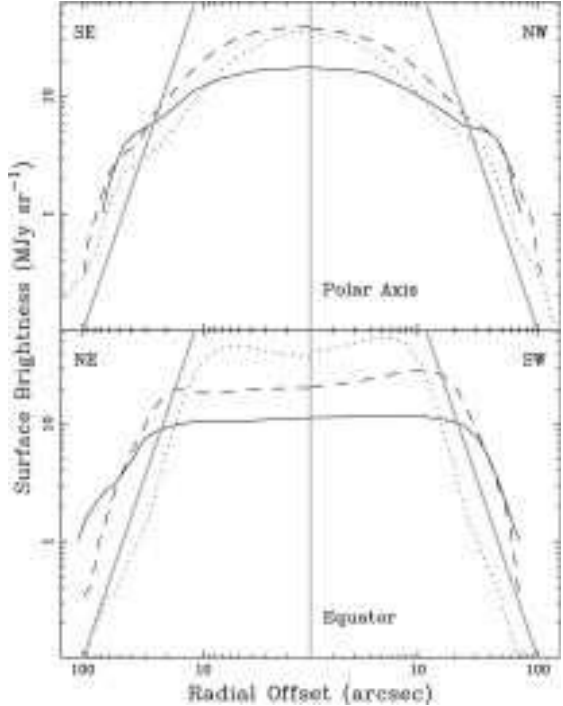} 
\end{center}
\figcaption{\label{logcuts}%
 Log-log plot of surface brightness profiles along the polar axis (top
 panel) and the equatorial plane (bottom panel).  The cuts are made
 along lines of $150\arcsec$ width on images that are convolved with the
 $40\arcsec$ beam to adjust the resolution. The direction of the cuts
 are indicated at the top corner of the panels.  The radial offsets are
 shown from $1\farcs28$ to $150\arcsec$ in each direction.  The slanted
 line in light gray is a line with the slope of $-3$ and is given to aid
 readers to read off the slope of the profile.} 
\end{figure}

In Figure \ref{logcuts}, we present log-log plots of surface brightness 
profiles along the bipolar axis (top panel) and the equatorial plane
(bottom panel).
These profiles are constructed from cuts of $150\arcsec$ width taken
in images that are convolved with the $40\arcsec$ beam to remove the
difference due to resolution.
Comparison between the top and bottom plots immediately indicates
distinct mass loss histories along these two directions.

The equatorial profiles follow a nearly perfect power-law with an index
of three along the edge of the nebula.
In an optically thin isothermal dust shell, a power-law density profile
of $\rho(r) \propto r^{-\alpha}$ will yield a power-law surface brightness
profile of $I(r) \propto r^{-(\alpha-1)}$.
Thus, the index of three we have observed in the equatorial profiles
suggests that the equatorial density profile follows $r^{-4}$.
This means that at the end of the AGB phase (presumably the most
effectively during the superwind phase) the rate of mass loss
was enhanced rapidly, since a constant mass loss at constant velocity 
yields a profile with an index of two.
Such a high power-law factor in density profiles of the late
AGB/superwind shell is consistent with recent reports of both
observational (e.g., \citealt{meixner04,h05}) and theoretical (e.g.,
\citealt{p04,s05}) studies.

Comparison between the equatorial profiles (Fig.\ \ref{logcuts}, top)
suggest possibly different mass loss histories along each edge of the
torus.  
While the southwestern profiles remain steep with an index of three,
the northeastern profiles are less steep especially beyond roughly
$50\arcsec$ in the $160\micron$.
Thus, there appears to be a larger amount of lower temperature dust
located at farther away from the central star on the northeast 
side of the torus (i.e., the $I_{160\micron}/I_{70\micron}$ ratio
becomes large as a function of radius along the northeastern edge).
One interpretation is that the period of the increased mass loss was
less intense yet prolonged on the northeastern edge than on the
southwestern edge. 
Such mass loss would result in a higher concentration of dust on the
southwestern edge of the torus, which would cause a stronger dust
emission as observed. 

The polar profiles (Fig.\ \ref{logcuts}, bottom) exhibit histories of
more complex mass loss. 
The outer part ($\age 50\arcsec$) appears very similar to
the equatorial profiles (showing a power-law with an index of three). 
However, the profiles flatten considerably (to an index of less than
0.5) roughly between $50\arcsec$ to $20\arcsec$ and then become steep 
again (to an index of about two) for $\ale 20\arcsec$.
Unfortunately, the possible rejevenation of mass loss in the inner part
is not entirely certain since the inner profiles are dominated 
by the bright torus of $40\arcsec$ thickness.
Nevertheless, the existence of the plateau region appears real.

Since the density profile follows $r^{-4}$ in the outer part of the
shell along both the polar and equatorial directions, it appears that 
the rate of mass loss was enhanced in all directions during the AGB
phase.  
From the data, we recognize this interesting history of mass loss in
this object, in which the initially isotropic mass loss enhancement
ensued along the equator towards the end of the AGB phase while it
ceased in the polar directions sometime before the end of the AGB
phase. 
This is the only apparent change in mass loss during the late AGB phase
prior to the formation of the equatorial density enhancement in the
shell. 
In fact, the observed surface brightness profiles suggests the presence
of a more prolate density distribution around the innermost toroidal
density distribution. 
This is consistent with what the layered shell model proposes for the
structure of the proto-PN shells (\citealt{ueta02,ueta03}) based on the
mid-IR and optical imaging surveys of the proto-PN shells
\citep{meixner99,ueta00,meixner02} as discussed above. 

Therefore, we conclude from these profiles that (1) mass loss was
enhanced nearly isotropically towards the end of the AGB phase, (2)
mass loss in the polar directions precipitously decreased prior to the
end of the AGB phase, and (3) the density distribution transformed from
prolate spheroidal to toroidal during the late AGB phase.
Although it is not entirely clear which parts of the profiles are
pertinent to the superwind epoch, it appears evident that the superwind
epoch at the end of the AGB phase is associated with a {\sl cessation}
of mass loss in the polar directions rather than an enhancement of it in
the equatorial directions.   
This is a strikingly contrasting view of the late AGB mass loss with
respect to a commonly-held assumption that there is an enhancement
of mass loss into the equatorial plane. 
Nevertheless, the present data provide another piece of evidence that
supports the presence of a transitional spheroidal shell between the
outer spherical AGB shell and the inner toroidal superwind shell as
proposed in the layered shell model \citep{ueta02}. 

A particularly important aspect of the present work is that this
transitional spheroidal shell has been observed in a bipolar PN. 
This is a direct observational evidence for the presence of the
spheroidal density distribution beyond the inner torus in a bipolar PN. 
It has been typically (and probably naively) assumed that there is an 
absence of matter along the polar axis in bipolar shells of evolved
stars, proto-PN shells and PNs alike.
However, the presence of matter along the polar axis is evident in this
bipolar PN, further increasing our confidence in the layered shell
model.  
The present data also demonstrate that one can probe the AGB mass loss
history via far-IR imaging of PNs as long as the shell structure is not
altered by the passage of the ionization front and a fast wind.

In the following we estimate some quantities concerning the remnant AGB
wind shell.
By adopting the distance of 1.2 kpc (e.g., \citealt{k98}), the peak
separation of $42\farcs3$ (at $160\micron$) translates into
$7.6\times10^{17}$ cm.
Given the spatial resolution effect, this value should be taken as a
lower limit for the inner diameter of the dust torus generated by the
AGB mass loss. 
If we adopt a typical expansion velocity of 20 km s$^{-1}$ (of a
superwind) and assume constant-velocity expansion, the dynamical age
of the dust torus is estimated to be $6.0 \times 10^{3}$ yr.
This value is consistent with theoretical predictions of the post-AGB
evolution into the PN phase of intermediate mass ($\age 3$ M$_{\odot}$)
stars (e.g., \citealt{b95}). 

\begin{figure}
\begin{center}
 \plotone{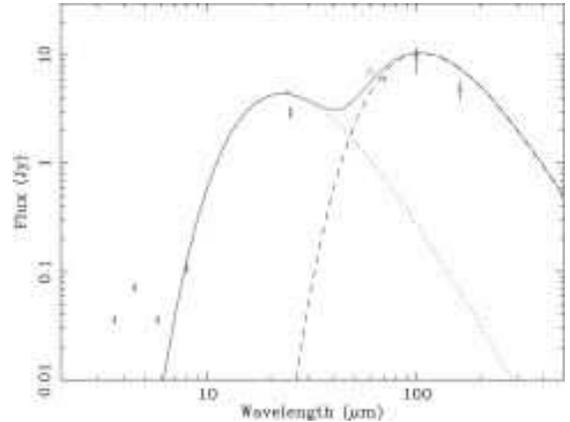} 
\end{center}
\figcaption{\label{sed}%
 The far-IR spectral energy distribution of NGC 650. Fluxes are taken
 from MIPS (square), IRAS (circle), and IRAC (triangle) observations.  
 The vertical bars indicate an error in each measurement.  The dashed
 line shows the best-fit to the colder AGB component of the dust shell
 (at 32 K), whereas the dotted line shows the higher temperature
 component (at 140 K).  The gray solid line indicates the total flux.} 
\end{figure}

Assuming that there is no significant line contamination in the measured
fluxes and that the dust shell is optically thin at far-IR, we can
estimate the dust temperature in the torus by fitting the far-IR SED of
the torus using the relation,   
\begin{equation}
 F_{\nu} = \Omega_{\rm beam} \tau_{\nu_{0}} 
  \left( \frac{\nu}{\nu_{0}} \right)^\beta B_{\nu}
 (T_{\rm dust}),
\end{equation}
where $F_{\nu}$ is flux density at frequency $\nu$, $\Omega_{\rm beam}$
is the instrumental beam area, $\tau (\nu) (= \tau_{\nu_{0}}
(\nu/\nu_{0})^\beta)$ is an optical depth at $\nu$ (scaled by a
power-law), and $T_{\rm dust}$ is the (representative) dust temperature
along the line of sight. 
Assuming fluxes longward of $100\micron$ represent the Rayleigh-Jeans
tail of the SED, the overall dust emissivity index can be fixed to
$1.6\pm0.5$.
By setting $\beta = 1.6$ in the above relation, we obtain the
least-squares fits of $T_{\rm dust}$ and $\tau(160\micron)$ under the
two-component dust assumption, using the present MIPS fluxes and the
past IRAS and IRAC fluxes. 
The dust temperatures are found to be $32\pm14$ K and $140\pm53$ K and
the optical depth at $160\micron$ is found to be $1.2 \times 10^{-5}$.
The SED and fitted curves are displayed in Figure \ref{sed}.
Given the power-law emissivity of the dust grains, the optical depth of
the shell in the optical (at V) is $0.1$: the torus is indeed optically
thin to visible light.

If the overall dust opacity, $\kappa$, is known, the measured far-IR
flux can be used to estimate the mass of dust in the torus through
\begin{equation}
 M_{\rm dust} \sim \frac{F_{\nu} D_{*}^2 c^2}{2 \nu^2 k T_{\rm dust}
  \kappa_\nu}. 
\end{equation}
Here, $D_{*}$ is the distance to NGC 650, $c$ is the speed of light, and
$k$ is Boltzmann constant.
\citet{semenov03} have computed Rosseland and Planck mean opacity
tables for the dusty medium based on the dust composition model for
accretion discs consisting of various silicates, organics (CHON
material), amorphous water ice, FeS, and iron having various particle
types.  
By adopting $\kappa = 0.4$ to 0.7 cm$^2$ g$^{-1}$ for the 
$\mbox{C/O} = 0.43$ and $\mbox{Fe}/(\mbox{Fe}+\mbox{Mg})=0.3$ (i.e.,
``normal'' iron abundance) case from their opacity table, the above
equation yields $1.4$ to $2.5 \times 10^{-2}$ M$_{\odot}$ for the
measured $160\micron$ flux of $4.83$ Jy and the estimated dust
temperature of $32$ K.  
With a canonical gas-to-dust ratio of 150, the total mass of the shell
is estimated to be 2 to 4 M$_{\odot}$.
The gas-to-dust ratio is highly uncertain, and so is the estimate of the
total mass of the shell.  
Nevertheless, the estimate is consistent with the inferred initial mass
of the central star ($\age 3$ M$_{\odot}$) based on the dynamical age of
the torus compared with theoretical post-AGB tracks (see above). 
Given the extent of the observed far-IR emission ($\sim 50\arcsec$) and
the inner radius of the torus ($\sim 20-30\arcsec$) plus the assumed
wind velocity of 20 km s$^{-1}$, the above values yield an overall mass
loss rate of $2 - 5 \times 10^{-4}$ M$_{\odot}$ yr$^{-1}$.

\subsection{The Post-AGB Shell}

We will now focus our attention to the inner, high-temperature torus
seen in the $24\micron$ band.
The $24\micron$ peaks do not simply represent the inside wall of the
dust torus which harbor high-temperature matter that is illuminated by
the central star.
If this were the case, the $24\micron$ peaks should have been spatially 
coincident with the 70 and $160\micron$ peaks.
Based on a spatial correlation between the 70 and $160\micron$ peaks and
the high ionization regions of \ion{He}{2} and [\ion{O}{3}], we have
concluded that the $24\micron$ emission is probably largely affected by
the [\ion{O}{4}] line at $25.9\micron$.

\begin{figure*}
 \begin{center}
 \includegraphics[width=\textwidth]{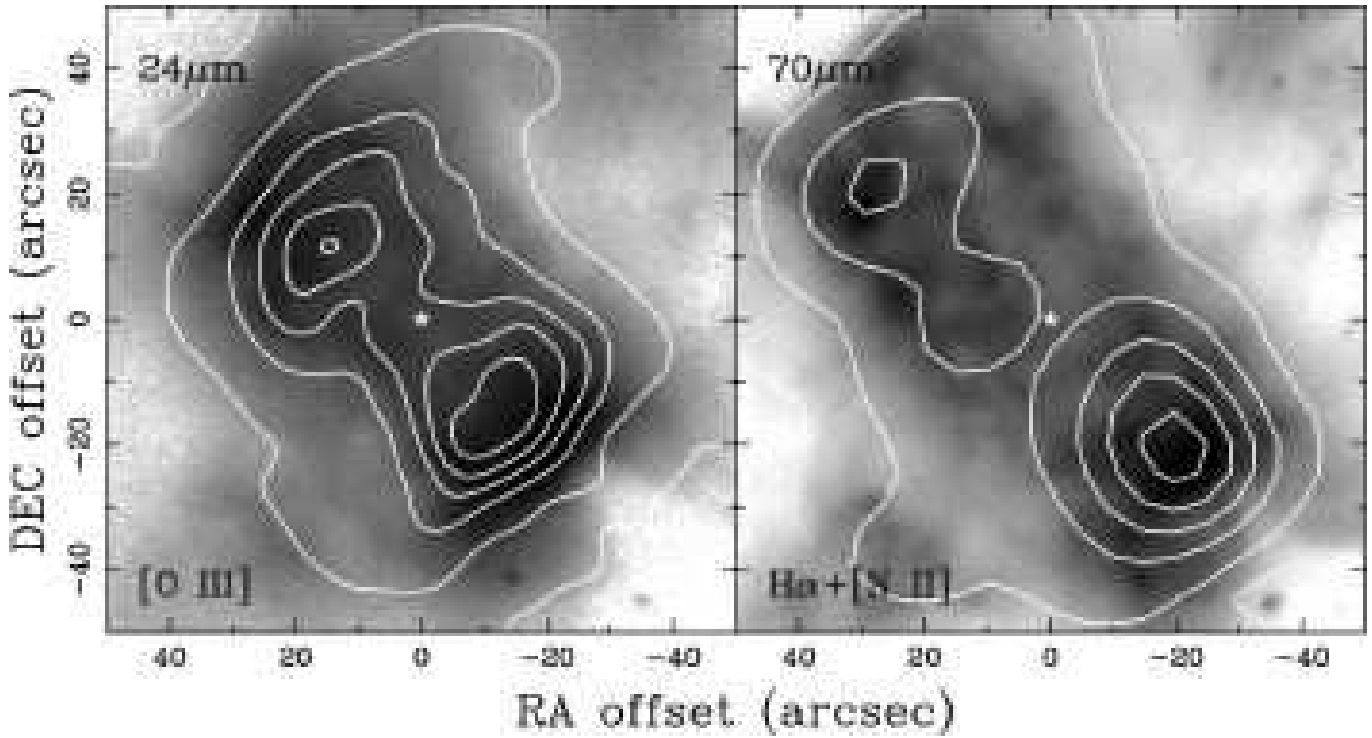}
 \end{center}
 \figcaption{\label{2x1map}%
 The deconvolved $24$ (left) and $70\micron$ (right) MIPS maps of the
 central $100\arcsec \times 100\arcsec$ region (N is up, E to the left) in
 contours overlaid with the IAC optical maps in the [\ion{O}{3}] (left)
 and H$\alpha$+[\ion{N}{2}] (right) bands in grayscale.
 The contours are, from top to bottom, 90 to 10\% of the peak with
 the 20\% interval.  The location of the central star ([0,0] offset) is
 indicated by a ``star'', showing the location of the $70\micron$
 central peak with respect to that of the central star.  The other
 display convention follows that of Fig.\ \ref{2x2map}.} 
\end{figure*}

It is not impossible, however, to have dust emission in such highly
ionized regions.
For example, the past Infrared Space Observatory observations detected
rather strong dust emission at $90\micron$ in the central region (i.e.,
the most ionized region) of the Helix nebula \citep{speck02}. 
The $70\micron$ map indicates that there is another emission peak near
the center of the nebula (Figure \ref{eqcuts}a; right panel of Figure
\ref{2x1map}).  
Even in the $160\micron$ map, there appears such emission at the nebula
center: the ``trough'' between the emission peaks is very shallow
(bottom panel of Figure \ref{2x2map}; Figure \ref{eqcuts}).

Aperture photometry of this central region (within the inner radius of
the low-temperature torus) yields 0.54 and 0.035 Jy respectively in the
70 and $160\micron$ bands.
These values are surprisingly close to the fluxes of the high-temperature
SED component at these wavelengths (0.51 and 0.031 Jy respectively at 70
and $160\micron$; Figure \ref{sed}), which cannot at all be entirely
attributed to photospheric emission of the central star ($3 \times
10^{-7}$ and $6 \times 10^{-8}$ Jy respectively at 70 and $160\micron$
extrapolated from the photospheric photometry in the IRAC bands
presented by \citealt{hora04}).  
Using the estimated far-IR fluxes from the inner torus and $\kappa \sim
0.2$ cm$^2$ g$^{-1}$ for dust of 140 K from the opacity table of
\citet{semenov03} in eqn (2), the mass of dust in this central peak is
$2.5 \times 10^{-7}$ M$_{\odot}$.
Given that only part of the measured $24\micron$ flux is dust-origin,
this mass estimate can only be a rough upper limit. 
The inner torus is thus very tenuous.

Figure \ref{2x1map} shows the central $100\arcsec \times 100\arcsec$
region of the deconvolved 24 and $70\micron$ contours overlaid
respectively with the [\ion{O}{3}] and H$\alpha$+[\ion{N}{2}]
grayscale maps.
The deconvolved $24\micron$ map shows an extension in the southwest peak
towards the center of the nebula.
This extension indicates the presence of some matter that fills the
inner cavity of the torus. 
The deconvolved $70\micron$ map also indicates the presence of some
dust in the inner cavity of the torus, which probably represents the
Rayleigh-Jeans tail of the high-temperature dust component as has been  
discussed above. 

Thus, the inner cavity of the torus appears to be filled with highly
ionized gas and some surviving dust behind the ionization front that 
has already engulfed the inner torus.
Where does this matter in the inner cavity come from?
There are two possibilities: this matter could come either from the
central star or back from the torus. 
If the cavity-filling matter comes from the star, the matter is likely
part of a fast wind.
Then, it is very peculiar that a fast wind has generated wind-blown
lobes along the bipolar axis, while maintaining a rather high
concentration of matter along the equatorial plane.

Alternatively, matter in the torus that is engulfed by the ionization
front can flow back towards the central star and fill the inner cavity
of the torus, if the pressure from a fast wind cannot hold it.
The back-flowing material from the torus initially fills the cavity, but
eventually disperses into the lobes.
The cavity-filling matter can get ionized by the radiation from the
central star to generate even O$^{+++}$.
In fact, \citep{bryce96} have observed multi-component [\ion{O}{3}] 5007
${\rm \AA}$ emission line profile in the central region of the nebula, and
the low-velocity components can be interpreted as back-flowing ionized
gas. 
Therefore, the inner torus represented by the $24\micron$ emission seems
to consist of the ionized portion of the torus behind the ionization
front, some ionized matter that has flown back towards the star filling
the inner cavity of the torus, and some surviving dust species.  
Thus, the presence of the post-AGB wind shell in NGC 650 is not very
likely. 

\section{Conclusions}

We have investigated the AGB and post-AGB mass loss history in a bipolar
PN, NGC 650, using far-IR maps at 24, 70, and $160\micron$ taken with
MIPS on-board the Spitzer Space Telescope. 
The far-IR surface brightness distribution shows two emission peaks at
all three MIPS bands.  
These peaks represent the limb-brightened edges of a near edge-on,
optically thin dusty torus.
Hence, this PN's far-IR structure is consistent with the known mid-IR 
structure of the PN progenitors, suggesting that a dusty torus in the
circumstellar shells of evolved stars endure the transition from the
post-AGB phase to the PN phase.
The measured fluxes above three $\sigma$ are $4.51\pm0.04$,
$6.04\pm0.30$, and $4.83\pm0.95$ Jy respectively at 24, 70, and
$160\micron$. 

While the 70 and $160\micron$ peaks are spatially coincident, the
$24\micron$ peaks are found interior to the peaks in the other bands.
Based on the spatial correlation with the distribution of [\ion{N}{2}],
H$\alpha$+[\ion{N}{2}], \ion{He}{2}, and [\ion{O}{3}] in the
optical, we have concluded that 
the far-IR toroidal structure consists of two components.
The 70 and $160\micron$ peaks represent low-temperature ($\sim 30$ K)
dust continuum emission from the remnant AGB wind shell. 
The $24\micron$ peaks correspond to the [\ion{O}{4}] line
emission at $25.9\micron$ arising from the highly ionized part of the
inner torus which has been engulfed by the ionization front and the
ionized matter that has flown back from the torus and filled the 
inner cavity of the torus.

The dynamical age of the remnant AGB wind shell is $6 \times 10^3$ yr,
assuming a constant expansion velocity of 20 km s$^{-1}$ and the
distance of 1.2 kpc. 
The least-squares fits to the far-IR SED have yielded the temperatures
of $32\pm14$ and $140\pm53$ K for the two-component dust shell system
and the optical depth of $\tau_{160\micron} = 1.2 \times 10^{-5}$
(i.e., $\tau_{\rm V} = 0.1$). 
The dust masses of the remnant AGB wind shell and the inner torus are
estimated to be 1.4 to $2.5 \times 10^{-2}$ and $2.5 \times
10^{-7}$ M$_{\odot}$, respectively.
For a canonical gas-to-dust ratio of 150, the total mass of the shell is 
2 to 4 M$_{\odot}$, yielding an overall mass loss rate of $2-5 \times
10^{-4}$ M$_{\odot}$ yr$^{-1}$.

From the far-IR surface brightness profiles of the nebula that show a
power-law index of as steep as three, we have also reconstructed this
PN's AGB mass loss history in which (1) mass loss was enhanced nearly
isotropically towards the end of the AGB phase, (2) the enhanced mass
loss in the equatorial directions ensued at the end of the AGB phase
while that in the polar directions precipitously ceased prior to the end
of the AGB phase, and (3) the density distribution transformed from
prolate spheroidal to toroidal during the late AGB phase.
Hence, the superwind phase at the end of the AGB phase is associated
with a {\sl cessation} of isotropically enhanced mass loss in the polar
directions rather than an enhancement of it only in the equatorial
directions.  

This is a strikingly contrasting view of the late AGB mass loss with
respect to a commonly-held assumption that there is an enhancement of
mass loss only along the equatorial plane. 
The present data thus provide another evidence for the presence of a
transitional spheroidal shell between the outer spherical AGB shell and
the inner toroidal superwind shell, further increasing our confidence in
the layered shell model for the circumstellar shells of post-AGB stars.
Moreover, the data indicate that the mass loss enhancement into the 
equatorial direction is not strictly uniform in all azimuthal angles. 
Mass loss appears to be less intense but more prolonged on the northeast
edge of the torus than the southwest edge, yielding higher concentration
of dust at the inner radius on the southwest edge. 
Nevertheless, the present data also demonstrate that one can probe the
AGB mass loss history via far-IR imaging of PNs as long as the shell
structure is not altered by the passage of the ionization front and a
fast wind. 

\acknowledgements
This research is based on archival data obtained with the Spitzer
Space Telescope, which is operated by the Jet Propulsion Laboratory,
California Institute of Technology under a contract with NASA. 
Support for this work was provided by a US National Research Council
Research Associateship Award and a NASA Postdoctoral Research Fellowship
Award.   
Additional support from the USRA SOFIA Office at NASA Ames Research
Center is also acknowledged.
The author is grateful to A.\ K.\ Speck and S.\ C.\ Casey for their
careful review and feedback to the initial draft of this paper, and to
the referee, B.\ Balick, for his comments that greatly improved the
final version of the paper.

\end{document}

%% file: tab1.tex
\begin{deluxetable}{lccc}
\tablecolumns{10} 
\tablewidth{0pt} 
\tablecaption{\label{flux}%
Far-IR Flux Measurements for NGC 650} 
\tablehead{%
\colhead{} &
\colhead{Flux} &
\colhead{Flux} &
\colhead{Flux} \\
\colhead{Measurements} &
\colhead{(Jy)} &
\colhead{(Jy)} &
\colhead{(Jy)}} 
\startdata 
 & $24\micron$ & $70\micron$ & $160\micron$ \\
\cline{1-4}\\
MIPS & $4.51\pm0.04$ & $6.04\pm0.30$ & $\phn4.83\pm0.95$ \\
\cline{1-4}\\
 & $25\micron$ & $60\micron$ & $100\micron$ \\
\cline{1-4}\\
IRAS PSC\tablenotemark{a} & $2.79\pm0.20$ & $6.80\pm1.43$ & $\phn9.29\pm0.51$ \\
IRAS FSC\tablenotemark{b} & $3.14\pm0.16$ & $7.28\pm0.66$ & $10.3\phn\pm0.9\phn$
\enddata
\tablenotetext{a}{IRAS Point Source Catalog data for IRAS 01391+5119}
\tablenotetext{b}{IRAS Faint Source Catalog data for IRAS F01391+5119}

\end{deluxetable}